\documentclass{article}

\newcommand{\be}{\begin{equation}}
\newcommand{\ee}{\end{equation}}
\newcommand{\ba}{\begin{eqnarray}}
\newcommand{\ea}{\end{eqnarray}}

\def\a{\alpha}

\def\e{\epsilon}

\def\vf{\varphi}
\def\g{\gamma}
\def\h{\eta}
\def\j{\psi}

\def\l{\lambda}
\def\m{\mu}
\def\n{\nu}

\def\p{\pi}
\def\q{\theta}

\def\x{\xi}
\def\z{\zeta}

\def\cm{{\cal M}}

\newcommand{\ov}{\overline}

\newcommand{\pr}{\prime}
\newcommand{\aand}{\;\;\;\mbox{and}\;\;\;}
\newcommand{\pa}{\partial}
\newcommand{\sx}{\sigma_x}
\newcommand{\sy}{\sigma_y}
\newcommand{\sz}{\sigma_z}

\def\sl#1{\rlap{\hbox{$\mskip 1 mu /$}}#1}

\def\I{\leavevmode\hbox{\small1\kern-3.8pt\normalsize1}}

\date{}

\title{{\bf Does the Higgs mechanism favour electron-electron bound states in 
Maxwell-Chern-Simons QED$_3$?}}

\author{Humberto Belich$^1$, Oswaldo M. Del Cima$^2$, \\
Manoel M. Ferreira Jr$^{1,3}$ and Jos\'e A. Helay\"el-Neto$^{1,2}$ \\
\small\it $^1$Centro Brasileiro de Pesquisas F\'\i sicas (CBPF),\\
\small\it Coordena\c c\~ao de Teoria de Campos e Part\'\i culas (CCP), \\
\small\it Rua Dr. Xavier Sigaud 150 - 22290-180 - Rio de Janeiro - RJ - Brazil. \\ 
\small\it $^2$Universidade Cat\'olica de Petr\'opolis (UCP), \\ 
\small\it Grupo de F\'\i sica Te\'orica (GFT), \\ 
\small\it Rua Bar\~ao do Amazonas 124 - 25685-070 - Petr\'opolis - RJ - Brazil. \\ 
\small\it $^3$Departamento de F\'\i sica,\\
\small\it Universidade Federal do Maranh\~ao (UFMA), \\
\small\it Campus Universit\'ario do Bacanga - 65085-580 - S\~ao Luiz - MA - Brazil.}

\begin{document}

\maketitle

\begin{abstract}
The low-energy electron-electron scattering potential is derived and discussed for the 
Maxwell-Chern-Simons model coupled to QED$_3$ with spontaneous symmetry breaking. One shows 
that the Higgs mechanism might favour electron-electron bound states.

\end{abstract}

\section{The Maxwell-Chern-Simons QED$_3$}

The action for the Maxwell-Chern-Simons model coupled to QED$_3$ \cite{djt} with a local 
$U(1)$-symmetry is given by\footnote{The metric is $\eta_{\m\n}$$=$$(+,-,-)$; $\m$,
$\n$$=$$(0,1,2)$, and the $\g$-matrices are taken as $\g^\m$$=$$(\sz,i\sx,-i\sy)$.}:
\ba
S_{\rm QED}=\int d^3x\biggl\{\!\!\!\!\!&-&\!\!\!\!\!\frac14 F^{\m\n}F_{\m\n}+i\ov\j \g^\m 
D_\m\j+
\frac12 \q\e^{\m\n\a}A_\m\pa_\n A_\a-m_e\ov\j \j + \nonumber\\
\!\!\!\!\!&-&\!\!\!\!\!y\ov\j \j\vf^*\vf+D^\m\vf^*D_\m\vf - V(\vf^*\vf)\biggr\}~, 
\label{action1}
\ea
where the $V(\vf^*\vf)$ is a sixth-power potential, being the most general renormalizable 
$U(1)$-invariant potential in three dimensions \cite{delcima}:
\be
V(\vf^*\vf)=\m^2\vf^*\vf+\frac\z2 (\vf^*\vf)^2+\frac\l3 (\vf^*\vf)^3~.
\ee
The covariant derivatives are defined as follows:
\be
D_\m\j=(\pa_\m+ieA_\m)\j \aand D_\m\vf=(\pa_\m+ieA_\m)\vf~.
\ee
In the action S$_{\rm QED}$, Eq.(\ref{action1}), $F_{\m\n}$ is the usual field strength for 
$A_\m$, $\j$
is a spinor field describing a fermion with positive spin polarization (spin up) and an 
anti-fermion with negative spin polarization (spin down) \cite{delcima},
whereas $\vf$ is a complex scalar field. In three space-time dimensions the positive- and 
negative-energy solutions have their polarization fixed by the signal of mass in the Dirac mass 
term \cite{binegar,delcima}. The mass dimensions of all the fields and parameters are displayed 
in the Table~\ref{table1}.
\begin{table}
\begin{center}
\begin{tabular}{|c||c|c|c|c|c|c|c|c|c|c|}
\hline
& $A_\mu$ & $\j$ & $\vf$ & $m_e$ & $\q$ & $e$ & $y$ & $\m$ & $\z$ & $\l$ \\
\hline
\hline
$d$ & $1/2$ & $1$ & $1/2$ & $1$ & $1$ & $1/2$ & $0$ & $1$ & $1$ & $0$ \\ 
\hline
\end{tabular}
\end{center}
\caption{Mass dimensions of the fields and parameters.}\label{table1}
\end{table}
The sixth-power potential is the responsible for breaking the electromagnetic
$U(1)$-symmetry. Analyzing the structure of the potential $V(\vf^*\vf)$, one must impose that 
it is bounded from below and it yields only
stable vacua (metastability is ruled out). These requirements reflect on the
following conditions on the parameters $\m$, $\z$ and $\l$ \cite{delcima}:
\be
\l>0~,~~\z<0 \aand \m^2\leq\frac{3\z^2}{16\l}~.
\ee

Considering $\langle\vf^*\vf\rangle=v^2$, the vacuum expectation value for the
scalar field product $\vf^*\vf$ is given by
\be
\langle\vf^*\vf\rangle=v^2=-\frac{\z}{2\l} + \sqrt{\left(\frac{\z}{2\l}\right)^2 - 
\frac{\m^2}{\l}}~,
\ee
while the minimum condition reads 
\be
\m^2+\z v^2+\l v^4=0~.
\ee

In order to preserve the manifest renormalizability of the model, one adopts the 't Hooft 
gauge:
\be
S_{{\rm R}_\x}=\int d^3x \biggl\{ -\frac{1}{2\x}(\pa^\m A_\m-\sqrt{2}\x M_A\q)^2\biggr\}~.
\ee
Then, by adding it up to the action (\ref{action1}), and assuming the following parametrization 
for the scalar field,  
\be
\vf=v+H+i\q~,
\ee 
where $H$ represents the Higgs scalar and $\q$ the would-be Goldstone boson, the 
Maxwell-Chern-Simons QED$_3$ action with the $U(1)$-symmetry spontaneously broken is as follows 
\ba
S_{\rm QED}^{\rm broken}=\int d^3x \biggl\{\!\!\!\!\!&-&\!\!\!\!\!\frac14 
F^{\m\n}F_{\m\n}+\frac12 M_A^2 A^\m A_\m + \frac12 \q\e^{\m\n\a} A_\m\pa_\n A_\a- 
\frac1{2\x}(\pa^\m A_\m)^2 + \nonumber\\
\!\!\!\!\!&+&\!\!\!\!\! \ov\j(i\g^\m D_\m - m)\j + \pa^\m H \pa_\m H + {\pa^\m}\q {\pa_\m}\q - 
\x M^2_A\q^2 + \nonumber\\
\!\!\!\!\!&-&\!\!\!\!\! y \ov\j \j (2vH + H^2+\q^2) + 2eA^\m(H{\pa_\m}\q - \q{\pa_\m}H) + 
\nonumber\\
\!\!\!\!\!&+&\!\!\!\!\! e^2 A^\m A_\m(2vH+H^2+\q^2) - \m^2((v+H)^2+\q^2) + \nonumber\\ 
\!\!\!\!\!&-&\!\!\!\!\! {\z\over2}((v+H)^2+\q^2)^2 - {\l\over3}((v+H)^2+\q^2)^3 
\biggr\}~,\label{action2}
\ea 
where the mass parameters $M^2_A$, $m$ and $M^2_H$, read
\be
M^2_A=2v^2e^2~,~~m=m_e+yv^2 \aand M^2_H=2v^2(\z + 2 \l v^2)~.
\ee

\section{Low-energy electron-electron scattering potential} 
The issue of electron-electron bound states in the Maxwell-Chern-Simons model coupled to planar 
QED has been addressed in the literature, since the end of the eighties 
\cite{kogan,girotti,hagen,dobroliubov}, motivated by possible applications to the 
parity-breaking superconductivity phenomenon.

In order to compute the scattering potential through the M{\o}ller electron-electron amplitude, 
we show the propagators associated to the Higgs ($H$), the fermion ($\j$) and the massive gauge 
boson ($A_\m$), which stem straightforwardly from the action (\ref{action2}), as presented 
below
\ba
\langle\ov\j(k)\j(k)\rangle\!\!\!\!\!&=&\!\!\!\!\!i\frac{{\sl k}+m}{k^2-m^2}~,~~
\langle H(k)H(-k)\rangle=\frac{i}2 \frac{1}{k^2-M_H^2} \aand \nonumber\\
\langle 
A_\m(k)A_\n(-k)\rangle\!\!\!\!\!&=&\!\!\!\!\!-i\biggl\{\frac{k^2-M_A^2}{(k^2-M_A^2)^2-k^2
\q^2}\biggl(\h_{\m\n}-\frac{k_\m k_\n}{k^2}\biggr)+\nonumber\\
\!\!\!\!\!&-&\!\!\!\!\!\frac{\x}{(k^2-\x M_A^2)}\frac{k_\m 
k_\n}{k^2}+\frac{\q}{(k^2-M_A^2)^2-k^2\q^2}i\e^{\m\a\n}k_\a\biggr\}~.
\ea

The non-relativistic scattering potential is nothing else than the two-dimensional Fourier 
transform of the lowest-order $\cm_{\rm total}$-matrix element:
\be
V(r)=\int \frac{d^2{\vec k}}{(2\p)^2}~\cm_{\rm total}^{\rm nr}~e^{i{\vec k}\cdot
{\vec r}}~.\label{fourier}
\ee
The $s$-channel amplitudes for the $e^-$-- $e^-$ scattering mediated by the Higgs and the gauge 
field are listed below:
\begin{enumerate}
\item Scattering amplitude by the Higgs:
\be
-i\cm_{e^-He^-}=\ov{u}(p_1)(2ivy)u(p^\pr_1)\langle H(k)H(-k)\rangle 
\ov{u}(p_2)(2ivy)u(p^\pr_2)~,\label{ehe}
\ee
\item Scattering amplitude by the massive gauge boson:
\be
-i\cm_{e^-Ae^-}=\ov{u}(p_1)(ie\g^\m)u(p^\pr_1)\langle A_\m(k)A_\n(-k)\rangle  
\ov{u}(p_2)(ie\g^\n)u(p^\pr_2)~,\label{eae}
\ee
\end{enumerate}
where $k^2$$=$$(p^\pr_1$$-$$p_1)^2$ is the invariant squared momentum transfer.

Now, bearing in mind that the non-relativistic $e^-$-- $e^-$ scattering potential in the Born 
approximation is obtained from the total scattering amplitude 
($\cm_{e^-He^-}$$+$$\cm_{e^-Ae^-}$) through the Fourier transform given by Eq.(\ref{fourier}), 
one gets:
\ba
V(r)=\!\!\!\!\!&-&\!\!\!\!\!{\frac{1}{2\p}} 2v^2y^2K_0(M_Hr) + \frac{e^2}{2\p}\biggl\{\left[C_+ 
- \frac{C}{m}M_+^2\right]K_0(M_+r) + \nonumber\\
\!\!\!\!\!&+&\!\!\!\!\! \left[C_- + \frac{C}{m}M_-^2\right]K_0(M_-r) + \nonumber\\ 
\!\!\!\!\!&+&\!\!\!\!\! 2\frac{l}{mr}C\left[M_+K_1(M_+r) - M_-K_1(M_-r)\right]\biggr\}~,
\label{potential}
\ea
where the positive definite constants $C_{+}$, $C_{-}$, $C$, and the squared masses $M^2_+$ and 
$M^2_-$, are given by:
\be
C_\pm=\frac12\biggl[1 \pm \frac{\q}{\sqrt{4M_A^2+\q^2}}\biggr]~,~~
C=\frac{1}{\sqrt{4M_A^2+\q^2}}~,
\ee
\be
M_\pm^2=\frac{1}{2}\left[2M_A^2+\q^2\pm |\q|\sqrt{4M_A^2+\q^2}\right]~,
\ee
with the mass poles $M^2_+$ and $M^2_-$ representing the two massive propagating quanta.

It should be stressed here that, by considering only the one-photon exchange diagrams in the 
non-relativistic limit, gauge invariance is spoiled \cite{hagen1}, therefore, two-photon 
exchange contributions have to be taken into account \cite{dobroliubov,szabo}. By adding up the 
two-photon contributions, so as to preserve gauge invariance, and the centrifugal barrier, to 
Eq.(\ref{potential}), the effective electron-electron scattering potential reads: 
\ba
V_{\rm eff}(r)=\!\!\!\!\!&-&\!\!\!\!\!{\frac{1}{2\p}} 2v^2y^2K_0(M_Hr) + \nonumber\\
\!\!\!\!\!&+&\!\!\!\!\!\frac{e^2}{2\p}\left\{\left[(C_+ - \frac{C}{m}M_+^2\right]K_0(M_{+}r) +
\left[C_- + \frac{C}{m}M_-^2\right]K_0(M_-r)\right\} + \nonumber\\
\!\!\!\!\!&+&\!\!\!\!\!\frac{1}{mr^2}\left\{l+\frac{e^2}{2\p}Cr[M_+K_1(M_+r) - 
M_-K_1(M_-r)]\right\}^2~, \label{effpotential}
\ea
where the term, in Eq.(\ref{effpotential}), proportional to $C^2$ arises from the two-photon 
exchange diagrams \cite{dobroliubov,szabo}.

\section{Conclusions}
The non-relativistic $e^-$-- $e^-$ scattering potential in the Born approximation for the 
Maxwell-Chern-Simons model coupled to QED$_3$ with spontaneous breaking of a $U(1)$-symmetry, 
given by Eq.(\ref{effpotential}), can be attractive provided a fine-tuning on the parameters is 
properly chosen.


\bigskip

\begin{thebibliography}{9}

\bibitem{djt}S. Deser, R. Jackiw and S. Templeton, Phys. Rev. Lett. {\bf 48}, 975 (1982) and 
Ann. Phys. (N.Y.) {\bf 140}, 372 (1982).

\bibitem{delcima} O.M. Del Cima, D.H.T. Franco, 
J.A. Helay\"el-Neto and O. Piguet, Phys. Lett. B {\bf 410}, 250 (1997) 
and Phys. Lett. B {\bf 416}, 402 (1998); M.A. De Andrade, O.M. Del Cima and J.A. 
Helay\"el-Neto, Il Nuovo Cimento {\bf 111}, 1145 (1998), and references therein.

\bibitem{binegar}B. Binegar, J. Math. Phys. {\bf 23}, 1511 (1982).

\bibitem{kogan}Ya.I. Kogan, JETP Lett. {\bf 49}, 225 (1989).

\bibitem{girotti}H.O. Girotti, M. Gomes, J.L. deLyra, R.S. Mendes, J.R.S. Nascimento and A.J. 
da Silva, Phys. Rev. Lett. {\bf 69}, 2623 (1992); H.O. Girotti, M. Gomes, A.J. da Silva, Phys. 
Lett. B {\bf 274}, 357 (1992).

\bibitem{hagen}C.R. Hagen, Phys. Rev. Lett. {\bf 71}, 202 (1993); H.O. Girotti, M. Gomes, J.L. 
deLyra, R.S. Mendes, J.R.S. Nascimento and A.J. da Silva, Phys. Rev. Lett. {\bf 71},
203 (1993).

\bibitem{dobroliubov}M.I. Dobroliubov, D. Eliezer, I.I. Kogan, G.W. Semenoff,
R.J. Szabo, Mod. Phys. Lett. A {\bf 8}, 2177 (1993).

\bibitem{hagen1}C.R. Hagen, Phys. Rev. D {\bf 31}, 848 (1985).

\bibitem{szabo}R.J. Szabo, I.I. Kogan and G.W. Semenoff, Nucl.
Phys. B {\bf 392}, 700 (1993). 

\end{thebibliography}
\end{document}